\documentclass{article}

\PassOptionsToPackage{numbers, compress}{natbib}

\usepackage[final]{neurips_2023}
\usepackage[utf8]{inputenc}
\usepackage[T1]{fontenc}
\usepackage{xurl}
\usepackage{hyperref}
\usepackage{enumitem}
\usepackage{booktabs}
\usepackage{amsfonts}
\usepackage{nicefrac}
\usepackage{microtype}
\usepackage{xcolor}
\usepackage{graphicx}
\usepackage{tabularx}
\usepackage{xcolor}

\title{Coordinated pausing: An evaluation-based coordination scheme for frontier AI developers}

\author{
  Jide Alaga\thanks{Jide Alaga worked on the project as part of the 2023 GovAI Winter Research Fellowship.} \\
  Centre for the Governance of AI \\
  \And
  Jonas Schuett \\
  Centre for the Governance of AI
}

\begin{document}

\maketitle

\begin{abstract}
As artificial intelligence (AI) models are scaled up, new capabilities can emerge unintentionally and unpredictably, some of which might be dangerous. In response, dangerous capabilities evaluations have emerged as a new risk assessment tool. But what should frontier AI developers do if sufficiently dangerous capabilities are in fact discovered? This paper focuses on one possible response: coordinated pausing. It proposes an evaluation-based coordination scheme that consists of five main steps: (1) Frontier AI models are evaluated for dangerous capabilities. (2) Whenever, and each time, a model fails a set of evaluations, the developer pauses certain research and development activities. (3) Other developers are notified whenever a model with dangerous capabilities has been discovered. They also pause related research and development activities. (4) The discovered capabilities are analyzed and adequate safety precautions are put in place. (5) Developers only resume their paused activities if certain safety thresholds are reached. The paper also discusses four concrete versions of that scheme. In the first version, pausing is completely voluntary and relies on public pressure on developers. In the second version, participating developers collectively agree to pause under certain conditions. In the third version, a single auditor evaluates models of multiple developers who agree to pause if any model fails a set of evaluations. In the fourth version, developers are legally required to run evaluations and pause if dangerous capabilities are discovered. Finally, the paper discusses the desirability and feasibility of our proposed coordination scheme. It concludes that coordinated pausing is a promising mechanism for tackling emerging risks from frontier AI models. However, a number of practical and legal obstacles need to be overcome, especially how to avoid violations of antitrust law.
\end{abstract}

\begin{figure}[ht!]
    \centering
    \includegraphics[width=\textwidth]{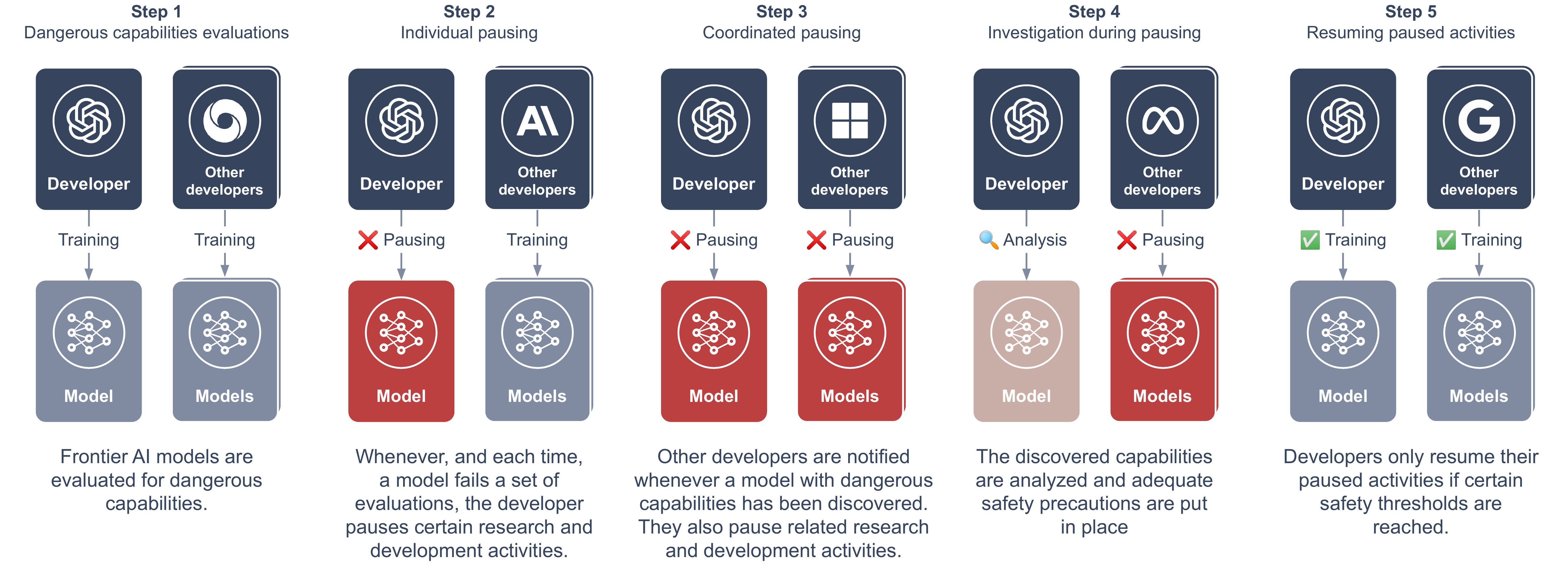}
    \caption{The main steps of our proposed evaluation-based coordination scheme}
    \label{figure_1}
\end{figure}

\section{Introduction}\label{1}

The past few years have shown a remarkable trend: more compute, larger datasets, and more parameters have led to the development of more capable artificial intelligence (AI) models. This phenomenon is commonly referred to as “scaling laws” \cite{Kaplan_et_al_2020, Hoffmann_et_al_2022, Sorscher_et_al_2022, Tay_et_al_2022, Caballero_et_al_2022, Villalobos_2023} and the claim that this trend will continue as the “scaling hypothesis” \cite{Gwern_2020}.\footnote[1]{Note that it has been argued that the current rate of scaling may be unsustainable \cite{Lohn_Musser_2022}.} While these scaling laws have been the driver of recent progress in AI development, they also have concerning implications. As models are scaled up, new capabilities can emerge unintentionally and unpredictably \cite{Ganguli_et_al_2022, Wei_et_al_2022}, some of which might be dangerous \cite{Shevlane_et_al_2023}.\footnote{Note that a recent paper expressed doubts about this phenomenon \cite{Schaeffer_Miranda_Koyejo_2023}.} For example, models might become able to persuade and manipulate people \cite{Park_et_al_2023, Meta_et_al_2022}, discover cyber vulnerabilities \cite{Lohn_Jackson_2022, Shevlane_et_al_2023}, or develop novel biological weapons \cite{Sandbrink_2023, Urbina_et_al_2022, Hendrycks_Mazeika_Woodside_2023}. These capabilities could be misused by malicious actors or used inadvertently by AI systems themselves. Some people even argue that certain combinations of capabilities could potentially lead to catastrophic outcomes \cite{Carlsmith_2022, Ngo_Chan_Mindermann_2022}.

In response, a suite of model evaluations that focus specifically on dangerous capabilities has emerged as a new risk assessment tool.\footnote{For an overview of other risk assessment techniques, see \cite{Koessler_Schuett_2023}.} In addition to developing these evaluations internally, some leading developers are taking proactive steps by involving external experts in safety evaluations before public releases. For example, before releasing GPT-4, OpenAI gave the Alignment Research Center’s evaluation team (ARC Evals) early access to the model to assess the extent to which it possessed dangerous capabilities \cite{OpenAI_2023a}. ARC Evals did the same with Anthropic’s Claude and Claude 2 \cite{Anthropic_2023a, Anthropic_2023b}. In both cases, ARC Evals concluded that the versions they tested did not have such dangerous capabilities \cite{ARC_Evals_2023a}. Yet, it remains unclear what developers should do if future evaluations actually discover sufficiently dangerous capabilities. This paper focuses on one possible response: coordinated pausing. The basic idea is that all frontier AI developers should pause certain research and development activities whenever and each time one of them discovers sufficiently dangerous capabilities. Developers only resume their paused activities if the discovered capabilities have been analyzed and adequate safety precautions have been put in place.

While there has been some work on evaluations for language models \cite{Chen_et_al_2021, Perez_et_al_2022, Liang_et_al_2022, Gehrmann_et_al_2022}, there is only limited work on dangerous capabilities evaluations. ARC Evals recently published a report in which they describe their methodology for assessing the capacity of language model agents to acquire resources, create copies of themselves, and adapt to novel challenges they encounter in the wild \cite{Kinniment_et_al_2023}. They have also published an update on their efforts to evaluate GPT-4 and Claude \cite{ARC_Evals_2023a}. Details of both efforts can be found in the GPT-4 system card \cite{OpenAI_2023a} and the Claude 2 model card \cite{Anthropic_2023b}. In addition to this work, there is only a single introductory paper on dangerous capabilities evaluations \cite{Shevlane_et_al_2023} and another paper that proposes a regulatory regime in which evaluations play a key role \cite{Anderljung_et_al_2023}.\footnote{Besides that, there only seem to be a few informal forum posts on the topic \cite{Krakovna_Shah_2023, Hubinger_2023, Christiano_2023}.}

Despite this shortage of literature, many experts take the topic very seriously. In a recent expert survey (\emph{N} = 51), 98\% of respondents somewhat or strongly agreed with the statement “AGI labs should run evaluations to assess their models’ dangerous capabilities”, while 93\% thought that “AGI labs should pause the development process if sufficiently dangerous capabilities are detected” \cite{Schuett_et_al_2023}. There have also been calls for a temporary moratorium in frontier AI development \cite{FLI_2023, Yudkowsky_2023}, but these calls were not linked to dangerous capabilities evaluations. Taken together, scholars and practitioners show considerable interest in evaluations, but the question of what should happen if sufficiently dangerous capabilities are in fact discovered remains underexplored.\footnote{Notably, ARC Evals recently announced plans to research responsible scaling policies, outlining how AI labs should scale, deploy, and contain models in the face of dangerous capabilities \cite{ARC_Evals_2023c}. Yet, this initiative remains an outlier, as there are few similar efforts in the broader AI community.} Against this background, the paper seeks to answer two research questions (RQs):

\begin{itemize}[leftmargin=2em]
    \item \textbf{RQ1:} How can frontier AI developers coordinate to pause if one of them discovers a model with sufficiently dangerous capabilities?
    \item \textbf{RQ2:} How desirable and feasible would an evaluation-based coordination scheme be?
\end{itemize}

The paper has three areas of focus. First, it focuses on \emph{dangerous capabilities evaluations}, but many considerations also apply to other types of evaluations (e.g. alignment evaluations).\footnote{In principle, this intervention can be implemented using any type of model evaluations related to catastrophic AI risk, such as alignment evaluations or evaluations for cooperative AI. However, comprehensively integrating other evaluations is beyond the scope of this paper. We leave this expansion for future work and focus here on existing methodologies.} Our vision for the proposed coordination scheme is that it should only be triggered if sufficiently dangerous capabilities are discovered. This is especially important given how intrusive the intervention is by nature. Second, the paper focuses on \emph{frontier AI developers}. We assume that the most concerning capabilities will only emerge in frontier AI models \cite{Anderljung_et_al_2023}, defined as “models that are both (a) close to, or exceeding, the average capabilities of the most capable existing models, and (b) different from other models, either in terms of scale, design (e.g. different architectures or alignment techniques), or their resulting mix of capabilities and behaviors” \cite{Shevlane_et_al_2023}. Third, the paper focuses on a \emph{collective solution}. The emphasis is not on what individual developers should do if they discover sufficiently dangerous capabilities, but how multiple (ideally all) frontier AI developers should respond to such a situation.

The paper proceeds as follows. Section \ref{2} proposes an evaluation-based coordination scheme. Section \ref{3} discusses four versions of that scheme. Sections \ref{4} and \ref{5} discuss the desirability and feasibility of coordinated pausing. Section \ref{6} concludes with suggestions for further research.

\section{An evaluation-based coordination scheme}\label{2}

In this section, we propose an evaluation-based coordination scheme for frontier AI developers. The scheme consists of five main steps as illustrated in Figure \ref{figure_1}:

\begin{itemize}[leftmargin=2em]
    \item \textbf{Step 1: Dangerous capabilities evaluations.} Frontier AI models are evaluated for dangerous capabilities (Section \ref{2.1}).
    \item \textbf{Step 2: Individual pausing.} Whenever, and each time, a model fails a set of evaluations, the developer pauses any further training and fine-tuning of that model. They also pause the development and deployment of similar models and do not publish related research (Section \ref{2.2}).
    \item \textbf{Step 3: Coordinated pausing.} Other developers are notified whenever a model with dangerous capabilities has been discovered. They also pause the development and deployment of similar models and do not publish related research (Section \ref{2.3}).
    \item \textbf{Step 4: Investigation during pausing.} The discovered capabilities are analyzed and adequate safety precautions are put in place (Section \ref{2.4}).
    \item \textbf{Step 5: Resuming paused activities.} Developers only resume their paused activities if certain safety thresholds are reached (Section \ref{2.5}).
\end{itemize}

In the following, we describe the five steps in more detail. For each of them, we identify key variables and list options. It is worth noting that, although the steps are described sequentially, there will be some overlap between them. For example, the investigation (Step 4) should arguably start as soon as dangerous capabilities are discovered (Step 2).

\subsection{Dangerous capabilities evaluations}\label{2.1}

\textbf{Step 1:} \emph{Frontier AI models are evaluated for dangerous capabilities.}

\textbf{Which models should be evaluated?} Our proposed coordination scheme should only apply to frontier AI models, as defined above. Frontier AI models are particularly risky because “(a) more capable models can excel at a wider range of tasks, which will unlock more opportunities to cause harm; and (b) novel models are less well-understood by the research community” \cite{Shevlane_et_al_2023}. Ideally, all such models should be evaluated for dangerous capabilities.\footnote{However, even if some developers do not run dangerous capabilities evaluations, under our proposed scheme they might still face significant public pressure to pause if it becomes widely known that dangerous capabilities have been discovered in another model and that a number of other developers have paused certain research activities in response.}

\textbf{What kind of evaluations?} Developers of frontier AI models need to run \emph{dangerous capabilities evaluations}. Shevlane et al. provide a good overview of different types of dangerous capabilities \cite{Shevlane_et_al_2023}. We have already mentioned the ability to persuade and manipulate people \cite{Park_et_al_2023, Meta_et_al_2022}, discover cyber vulnerabilities \cite{Lohn_Jackson_2022, Shevlane_et_al_2023}, and develop novel biological weapons \cite{Sandbrink_2023, Urbina_et_al_2022, Hendrycks_Mazeika_Woodside_2023}. Other potentially dangerous capabilities might include situational awareness, i.e. a model’s ability to to refer to and make predictions about itself as distinct from the rest of the world \cite{Cotra_2022, Ngo_Chan_Mindermann_2022}; power-seeking behavior, i.e. active efforts by a model to gain and maintain power in ways that its developers did not intend \cite{Carlsmith_2022, Turner_Tadepalli_2022, Turner_et_al_2023, Krakovna_Kramar_2023}; and long-horizon planning, i.e. a model’s ability to make sequential plans that involve multiple steps, unfolding over long time horizons \cite{Shevlane_et_al_2023}. We are aware of evaluations for power-seeking behavior \cite{ARC_Evals_2023a, Kinniment_et_al_2023} and efforts to develop evaluations for deception \cite{Apollo_Research_2023}, situational awareness \cite{Evans_2023}, and manipulation \cite{AI_Objectives_Institute_2023}. We are unaware of evaluations for other capabilities, such as the ability to exploit vulnerabilities in software systems or develop weapons.\footnote{But note that the red team OpenAI commissioned before releasing GPT-4 assessed the model’s ability to discover and exploit cybersecurity vulnerability, and support social engineering. The red team also tested whether GPT-4 could provide the necessary information to develop, acquire, or disperse nuclear, radiological, biological, and chemical weapons \cite{OpenAI_2023a}.} In any case, our proposed coordination scheme should only apply to evaluations that try to discover \emph{sufficiently} dangerous capabilities, which should be interpreted restrictively. It is beyond the scope of this paper to suggest what these danger thresholds should be. However, the intervention we are proposing is very intrusive and may only be appropriate if \emph{ex post} remedies would be insufficient.

\textbf{Who creates, maintains, and runs the evaluations?} Dangerous capabilities evaluations could be created and maintained by the developers themselves. A number of frontier AI developers already seem to have internal evaluation programs \cite{Shevlane_et_al_2023, Krakovna_Shah_2023}, though it is difficult to comment on such efforts from the outside. Alternatively, third-party organizations like ARC Evals or Apollo Research or academic centers like Stanford's Center for Research on Foundation Models could take on these roles, either on their own or in collaboration with developers. The task of running evaluations could also be shared between developers and third parties. Another option is having one developer scrutinize another's work, although a recent survey found little support for inter-lab scrutiny \cite{Schuett_et_al_2023}. It is worth noting that the actor who creates and maintains the evaluations does not necessarily need to be the one who runs them. For example, evaluations might be created by a third party, but run by the developers themselves.

\textbf{How is compliance monitored and enforced?} Developers are incentivized to deploy models quickly, but running evaluations takes time. We should therefore expect that some developers will not run dangerous capabilities evaluations. This raises the questions of how compliance should be monitored and enforced. By default, there are no monitoring and enforcement mechanisms. The scheme would rely on the goodwill of frontier developers. However, developers could support various monitoring mechanisms on a voluntary basis. One option would be to create and maintain a whistleblower program. Employees who find out that their company decided against running evaluations on frontier models could reveal such information to a trusted party, such as an ethics board \cite{Schuett_Reuel_Carlier_2023}, an internal audit team \cite{Schuett_2023}, or a regulator. Another option would be to give a third party certain investigative powers (e.g. the right to access documents, interview employees, or attend meetings). This could include an ethics board \cite{Schuett_Reuel_Carlier_2023}, an auditor (e.g. ARC Evals), an industry body (e.g. Frontier Model Forum), a multi-stakeholder organization (e.g. Partnership on AI), or a regulator. If developers make legally binding commitments to run dangerous capabilities evaluations, they might face contractual liability if they break them. Finally, if at some point frontier AI developers are required by law to run dangerous capabilities evaluations, such laws would likely entail provisions about the enforcement of such requirements (e.g. via fines and penalties).

\subsection{Individual pausing}\label{2.2}

\textbf{Step 2:} \emph{Whenever, and each time, a model fails a set of evaluations, the developer pauses further training and fine-tuning of that model. They also pause the development and deployment of similar models and do not publish related research.}

\textbf{When do developers pause?} Developers should pause certain research and development activities whenever, and each time, a model fails a set of dangerous capabilities evaluations. This trigger is one of the main differences to a general moratorium on frontier AI development \cite{FLI_2023}. But when exactly does a model “fail” a set of evaluations? Defining this danger threshold is one of the most important parts of our proposed scheme. Unfortunately, it is also one of the most difficult parts. On the one hand, the threshold should be very high. Since pausing is a very intrusive intervention, it should only be triggered in rare cases. On the other hand, the pausing scheme is intended to prevent severe harm. It is therefore crucial to avoid false negatives, i.e. models that do have dangerous capabilities do not trigger a pause. How to balance these and other considerations is still an open question. It is beyond the scope of this paper to suggest danger thresholds or tiers for evaluation results.

\textbf{What do they pause?} Developers should pause four types of activities:

\begin{itemize}[leftmargin=2em]
    \item \textbf{Development.} First, they should pause the development of the model that failed the evaluations. More precisely, they should pause any ongoing training runs and delay any scheduled training runs \cite{Shevlane_et_al_2023}.\footnote{Shevlane et al. note that frontier AI developers should factor in potential pauses in their research plans \cite{Shevlane_et_al_2023}. For example, they should plan in advance how they would backfill vacant computing resources with other projects. They should also avoid promising certain release dates.} This also applies to fine-tuning and reinforcement learning from human feedback (RLHF). To avoid situations in which the same or similar dangerous capabilities emerge in other models, developers should also pause the development of similar models. Models can be similar in terms of their architecture, size, training data, or compute, to name just a few criteria. It is beyond the scope of this paper to suggest measures and thresholds for similarity. This is an open question that needs more research. 
    \item \textbf{Deployment.} Second, developers should pause the deployment of the model that failed the evaluations. It would be inconsistent if developers were required to pause the development process, but still allowed to deploy the model. For the same reasons, they should also pause the deployment of similar models. 
    \item \textbf{Access.} Third, developers should restrict access to similar models that they have already deployed \cite{OBrien_Williams_Ee_forthcoming}. This is not possible if such models are open-sourced. Participating developers should therefore not open-source frontier models \cite{Seger_et_al_forthcoming} and instead deploy them via an API \cite{Shevlane_2022, Solaiman_2023}. We wish to emphasize that open-sourcing non-frontier models—and the vast majority of models are non-frontier models—is often a valuable contribution to the AI research community and society more generally. 
    \item \textbf{Related research.} Fourth, developers should pause the publication of related research. This avoids the possibility of other actors quickly developing models with similar capabilities. The research necessary to develop such models should therefore not be publicly available (sometimes this is already the case). Relatedly, developers should arguably also pause doing related research itself, but we are less certain about that.
\end{itemize}

\textbf{How is compliance monitored and enforced?} The monitoring and enforcement options are similar to the ones mentioned above (Section \ref{2.1}). By default, individual pausing cannot be enforced, but developers could take voluntary steps to support the monitoring of compliance (e.g. by maintaining a whistleblower program). They could also give a third party certain investigative powers (e.g. an ethics board or an auditor). If developers are legally required to pause, supervisory authorities would likely be able to monitor and enforce compliance.

\subsection{Coordinated pausing}\label{2.3}

\textbf{Step 3:} \emph{Other developers are notified whenever a model with dangerous capabilities has been discovered. They also pause the development and deployment of similar models and do not publish related research.}

\textbf{How are other developers notified?} There are three ways in which other developers can be notified. First, the developer who has trained the model with dangerous capabilities could notify other developers directly (e.g. via email). Second, the developer could make the incident public. For example, they could tweet about the incident, publish a blog post (e.g. similar to OpenAI \cite{OpenAI_2023c}), or make an entry in an incident database \cite{McGregor_2021}. Third, the developer could notify a third party who could then notify other developers. The third party could be a mutual auditor: a single organization runs evaluations on frontier models of multiple developers and notifies all developers they work with if a model fails a set of evaluations. The auditor’s right to notify other developers (and the developers’ right to be notified) would have to be specified in the contract between the auditor and each of the participating developers (e.g. it could be a standard clause in the audit agreement). Other potential third parties include industry bodies (e.g. Frontier Model Forum), multi-stakeholder organizations (e.g. Partnership on AI), or regulators, especially if they were involved in previous steps. Involving a third party could be an elegant way to avoid some of the antitrust concerns mentioned in Section \ref{5}.

\textbf{What do they pause?} Other developers should pause the development and deployment of models similar to the one with dangerous capabilities, restrict access to similar models that have already been deployed, and not publish related research (Section \ref{2.2}). Again, it is beyond the scope of this paper to define which models are “similar” and which research is “related”.

\textbf{How is compliance monitored and enforced?} The monitoring and enforcement options are similar to those mentioned above (Section \ref{2.1} and \ref{2.2}).

\subsection{Investigation during pausing}\label{2.4}

\textbf{Step 4:} \emph{The discovered capabilities are analyzed and adequate safety precautions are put in place.}

\textbf{What happens during pausing?} During pausing, four things should happen. First, the model should be contained (e.g. via boxing or air-gapping) as soon as dangerous capabilities are discovered \cite{Babcock_et_al_2019, ARC_Evals_2023a, Schuett_et_al_2023}. Developers should also increase their efforts to prevent leakage and theft of the model. In the future, this might require military-grade information security sufficient to defend against nation states \cite{Schuett_et_al_2023}. Second, the model should be analyzed to determine why it failed the evaluations. This might involve additional tests of the model’s behavior or attempts to understand the inner workings of the model via interpretability research.\footnote{Although there are promising developments \cite{Nanda_et_al_2023}, the field of mechanistic interpretability is still in its infancy \cite{Nanda_2022, Olah_2023}. Conducting interpretability research is very time-consuming and does not yet seem practical in a pausing context.} Third, the developer should take measures to make the model safer, for example, via fine-tuning \cite{Solaiman_Dennison_2021}, reinforcement learning from human feedback (RLHF) \cite{Christiano_et_al_2017, Ziegler_et_al_2019, Lampert_et_al_2022}, or reinforcement learning from AI feedback (RLAIF), more commonly known as “constitutional AI” \cite{Bai_et_al_2022}---though it is also possible that existing techniques will not be sufficient. Fourth, the developer should put in place adequate safety controls. For example, they might only deploy the model in stages \cite{Solaiman_et_al_2019, Solaiman_2023}, via an API \cite{Solaiman_et_al_2019, Cohere_OpenAI_AI21_2022, Shevlane_2022, Solaiman_2023}, and with certain restrictions (e.g. who can use the model, how they can use the model, and whether the model can access the internet). Shevlane et al. list a number of variables that affect the risk level of deployment \cite{Shevlane_et_al_2023}.

\textbf{Who does the investigation?} Most of the above-mentioned activities need to be performed by the developer of the model that has triggered the pause (e.g. model containment). However, the developer could also be supported by other actors. For example, if the evaluations are run by an independent auditor, this auditor will often be best equipped to analyze why the model has failed the evaluations. The developer might also bring in additional auditors (e.g. to replicate the evaluation results or run additional evaluations) or external researchers (e.g. to conduct interpretability research). In theory, it would also be conceivable that other developers support the investigations, but in practice it does not seem politically feasible (e.g. because of antitrust and confidentiality concerns).\footnote{In a recent expert survey (\emph{N} = 51), inter-lab scrutiny was one of the least supported items, though it still received more agreement than disagreement \cite{Schuett_et_al_2023}. On a scale from -2.0 (strongly disagree) to 2.0 (strongly agree), the medium (\emph{M}) rating for inter-lab scrutiny was 0.7. It is worth noting that, while not statistically significant, we saw higher support for this statement from respondents from AGI labs (\emph{M} = 1.2) in comparison to respondents from academia (\emph{M} = 0.3) and civil society (\emph{M} = 0.2).} Other developers should also take corresponding measures where appropriate (e.g. taking additional measures to make their models safer and strengthening their safety controls). The entire investigation should probably be overseen by a third party (e.g. an auditor, ethics board, multi-stakeholder organization, or regulator).

\subsection{Resuming paused activities}\label{2.5}

\textbf{Step 5:} \emph{Developers only resume their paused activities if certain safety thresholds are reached.}

\textbf{When can developers resume their paused activities?} The decision to resume the paused activities raises some of the same issues as the initial decision to pause. Defining danger thresholds (when should frontier AI developers pause?) and safety thresholds (when can they resume their paused activities?) are essentially two sides of the same coin. However, it might make sense to set the safety threshold higher than the danger threshold. To reach that threshold, the model may need to pass a more demanding and more diverse set of evaluations. As above (Section \ref{2.2}), more research is needed to determine when a developer’s understanding of certain capabilities is sufficient and what kind of safety precautions are adequate. This will likely require a holistic evaluation of each case, taking into account technical, social, and organizational factors. The decision will inherently involve uncertainties. And the actor who makes this decision will likely need discretion.

\textbf{Who decides when safety thresholds are reached?} Each developer could make that decision for themselves. Absent legal requirements or voluntary commitments, other developers will likely want to resume their paused activities before the developer of the model with dangerous capabilities. It would also be conceivable that all participating developers make a collective decision (e.g. they could vote), but this may raise antitrust concerns. Another option would be that a third party makes that decision on behalf of the developers. This could be the auditor who ran the evaluations that discovered the dangerous capabilities, an industry body (e.g. Frontier Model Forum), a multi-stakeholder organization (e.g. Partnership on AI), or a regulator.

In this section, we have described the five steps of our proposed coordination scheme. We have identified key variables and listed options. However, we have not discussed how different options could be combined in a coherent way. We will turn to this next.

\section{Concrete versions of the proposed coordination scheme}\label{3}

In this section, we discuss four concrete versions of the coordination scheme proposed above (Section \ref{2}). In the first version, pausing is completely voluntary and relies on public pressure on developers (Section \ref{3.1}). In the second version, participating developers collectively agree to pause under certain conditions (Section \ref{3.2}). In the third version, a single auditor runs evaluations on models of multiple developers and they agree to pause if any model fails a set of evaluations (Section \ref{3.3}). In the fourth version, developers are legally required to run evaluations and pause if dangerous capabilities are discovered (Section \ref{3.4}). For each of the four versions, we explain how they work, suggest variations, discuss their main benefits and limitations, and make recommendations. Table \ref{table_1} contains an overview of the four versions.

\renewcommand{\arraystretch}{1.7}
\begin{table}[t!]
  \centering
  \scriptsize
  \begin{tabularx}{\textwidth}{>{\raggedright}p{0.2\linewidth}>{\raggedright}p{0.17\linewidth}>{\raggedright}p{0.17\linewidth}>{\raggedright}p{0.17\linewidth}>{\raggedright\arraybackslash}p{0.16\linewidth}}
    \toprule
        & \textbf{Voluntary pausing} & \textbf{Pausing agreement} & \textbf{Mutual auditor} & \textbf{Required pausing} \\
    \midrule
        \textbf{Step 1: Dangerous capabilities evaluations} & & & & \\
        Which models should be evaluated? & Frontier AI models & --- & --- & --- \\
        What kind of evaluations? & Dangerous capabilities evaluations & --- & --- & --- \\
        Who creates, maintains, and runs the evaluations? & Developers and/or third party & Auditor & Auditor & Auditor \\
        How is compliance monitored and enforced? & No monitoring and enforcement, only public pressure and whistleblowing & Other developers & Auditor & Regulator \\
        \textbf{Step 2: Individual pausing} & & & & \\
        When do developers pause? & A frontier AI model fails a set of dangerous capabilities evaluations & --- & --- & --- \\
        What do they pause? & Development, deployment, related research & --- & --- & --- \\
        How is compliance monitored and enforced? & No monitoring and enforcement, only public pressure and whistleblowing & Other developers & Auditor & Regulator \\
        \textbf{Step 3: Coordinated pausing} & & & & \\
        How are other developers notified? & Results of evaluations and incidents are made public & Developer & Auditor & Regulator \\
        What do they pause? & Development, deployment, related research & --- & --- & --- \\
        How is compliance monitored and enforced? & No monitoring and enforcement, only public pressure and whistleblowing & Other developers, contractual penalties & Auditor & Regulator \\
        \textbf{Step 4: Investigation during pausing} & & & & \\
        What happens during pausing? & Model is contained, incident is analyzed, model is made safer, safety controls are implemented & --- & --- & --- \\
        Who does the investigation? & Developer and/or third party & Developer and/or third party & Auditor, developers cooperate & Auditor, supervised by regulator, developers cooperate \\
        \textbf{Step 5: Resuming paused activities} & & & & \\
        When can developers resume their paused activities? & Safety threshold is reached & --- & --- & --- \\
        Who decides when this is the case? & Developers (individually) & Developers (collectively) & Auditor & Regulator \\
    \bottomrule
  \end{tabularx}
  \vspace{0.7em}
  \caption{Overview of four concrete versions of the proposed coordination scheme} \label{table_1}
  \vspace{-2em}
\end{table}

\subsection{Voluntary pausing}\label{3.1}

In the first version, pausing is completely voluntary and relies on public pressure on developers.

\textbf{How it works.} Frontier AI developers do not make any commitments to pause and there are no legal requirements. Running evaluations and pausing is completely voluntary, but developers face public pressure to do so. In particular, they are expected to publish the results of dangerous capabilities evaluations (e.g. in short summary reports) before deploying frontier models or publishing related research. Some developers have already made a high-level commitment along these lines \cite{The_White_House_2023b}. These evaluations may be conducted by external auditors or developers themselves. Publishing evaluation results creates a “public commentary period” which allows the wider AI research community to scrutinize the evaluation results and raise concerns. Depending on how serious such concerns are, the developer of that model and other developers might be pressured to pause certain research and development activities. The length and nature of the pausing period would be at the discretion of the developers, but it would be affected by how much pressure is put on them. Monitoring relies on whistleblowing, though in individual cases, regulators may request additional information \cite{Zakrzewski_2023}.

\textbf{Variations.} In the version described above, developers do not make any commitments. But it would be conceivable that they make at least a soft commitment. For example, they could publish a blog post in which they commit to evaluate frontier models and pause if dangerous capabilities are discovered in any frontier model. Google DeepMind’s post on dangerous capabilities evaluations is promising sign \cite{Shevlane_2023}, but it remains vague. The post does not specify what kinds of evaluations Google DeepMind currently runs, it does not contain an explicit commitment to pause, and it does not define any thresholds.

\textbf{Benefits.} Of the four versions, voluntary pausing is the most feasible one. It is close to the status quo. Frontier AI developers already run evaluations and there is already an expectation to pause if dangerous capabilities are discovered \cite{Shevlane_et_al_2023, Schuett_et_al_2023}, though this pressure might not yet be strong enough. A benefit of this version is that it is fairly light-touch. Developers do not have to negotiate a contract and policymakers do not have to pass new laws or regulations. The administrative burden on developers is also comparably small. Another benefit of this version is that it is fairly flexible. Since the appropriate response to a set of failed evaluations is not enshrined in any way, this version can quickly react to scenarios in which frontier models are less dangerous than expected (do not pause) or even more dangerous (take more extreme measures). Changing expectations often takes less time than amending a contract, regulation, or law. A single incident might be sufficient to cause a public outcry that puts significant pressure on developers \cite{Cihon_et_al_2021}.

\textbf{Limitations.} From a societal perspective, voluntary pausing is not particularly desirable. Since there are no monitoring and enforcement mechanisms, there is no reliable way to ensure compliance. While public pressure does incentivize compliance to some extent, other incentives might be even stronger (e.g. money, prestige, national interests), especially if the stakes are high. Developers might also be reluctant to publish the results of their evaluations. They would open themselves up to public scrutiny with unpredictable PR risks. And even if they do publish the results of their evaluations, the public commentary period might still lead to counterproductive outcomes. For example, one could imagine that the discourse becomes politicized and dominated by non-safety considerations. Another limitation is that there would be little consistency between developers in terms of evaluations and pausing. While this would still be better than the status quo, a single developer who does not participate might be enough to cause severe harm. Finally, since models are not tested by an independent third party, developers can—intentionally or not—run evaluations in a way that ensures their models remain below the danger threshold. This concern is related to Goodhart's law which states that “when a measure becomes a target, it ceases to be a good measure” \cite{Strathern_1997} (Section \ref{5}).

\textbf{Recommendation.} We would strictly prefer any of the other versions over this one. But as long as there are no pausing agreements (Section \ref{3.2}), audit agreements (Section \ref{3.3}), or pausing requirements (Section \ref{3.4}), external stakeholders (e.g. independent researchers and civil society organizations) should put pressure on frontier AI developers to run evaluations and pause if sufficiently dangerous capabilities are discovered. In particular, they should voice their expectations that developers publish the results of evaluations to allow for a “public commentary period”. They should also advocate for binding commitments and eventually legal requirements. Overall, this version should only be seen as an intermediate solution. Once public pressure is strong enough, other versions will likely become more feasible.

\subsection{Pausing agreement}\label{3.2}

In the second version, participating developers collectively agree to pause under certain conditions.

\begin{figure}
    \centering
    \includegraphics[width=0.8\textwidth]{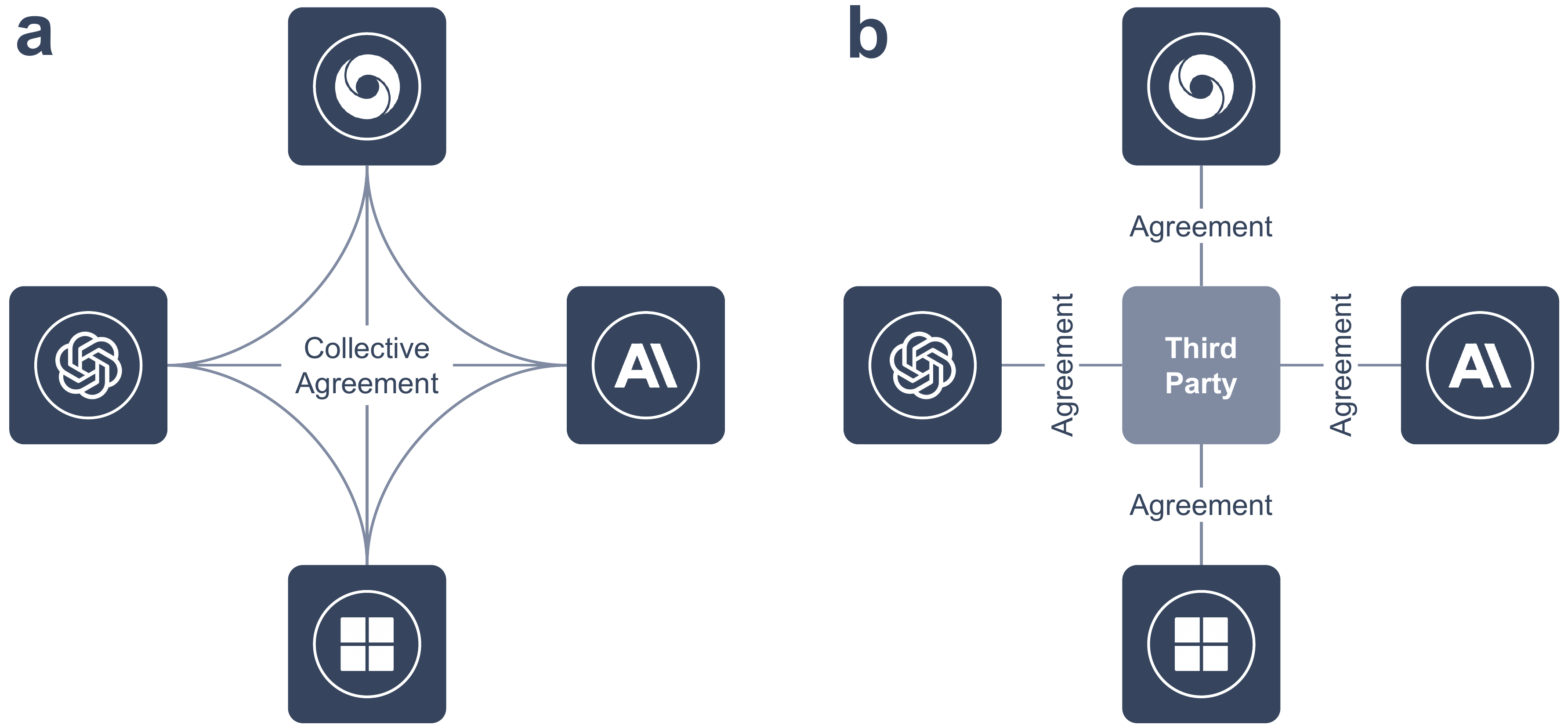}
    \caption{Collective pausing agreement \textbf{(a)} and individual pausing agreements with a third party \textbf{(b)}}
    \label{figure_2}
\end{figure}

\textbf{How it works.} Participating developers negotiate a contract (Figure \ref{figure_2}a). In that contract, they all commit to commission a third party to run dangerous capabilities evaluations, notify the other contracting parties if a model fails a set of evaluations, and pause certain research and development activities until certain safety thresholds are reached. Compliance is monitored by the developers themselves and enforced via contractual penalties. Conflicts resulting from the agreement are resolved by an independent arbitrator (e.g. a panel of experts).

\textbf{Variation.} Instead of a collective pausing agreement, developers could make individual agreements with a third party (Figure \ref{figure_2}b). An obvious candidate would be the Frontier Model Forum, an industry body founded by Anthropic, Google DeepMind, Microsoft, and OpenAI in July 2023 \cite{OpenAI_2023b}. As a condition of membership, the Frontier Model Forum could legally require developers to commit to coordinated pausing in their terms and conditions. Specifically, it could mandate that members notify the Forum whenever one of their models fails a set of dangerous capabilities evaluations, and agree to pause development when notified that any member's model has failed evaluations. While the Forum would not conduct evaluations itself, by making coordinated pausing a binding membership requirement, it could serve as an intermediary to facilitate implementation of the intervention. This approach is somewhat similar to using a mutual auditor (Section \ref{3.3}). The main difference is that the Forum would leverage membership rules rather than direct auditing agreements to coordinate pausing, while preserving antitrust law compliance.

\textbf{Benefits.} The main benefit of a pausing agreement over voluntary pausing (Section \ref{3.1}) is that participating developers make a legally binding commitment. Compliance with this commitment is monitored and enforced which will likely lead to higher degrees of compliance. It also seems more realistic that developers enter into a pausing agreement than that policymakers create pausing requirements (Section \ref{3.4}),\footnote{Collective or individual pausing agreements seem similarly feasible as an agreement with a mutual auditor (Section \ref{3.3}).} though the regulatory debate in the US and UK has picked up speed.

\textbf{Limitations.} Some scholars and practitioners have voiced the concern that this kind of cooperation between developers violates US and EU antitrust laws. We imagine that individual agreements with a third party (e.g. the Frontier Model Forum) would not run into this problem. However, since we are not antitrust experts and a legal analysis is beyond the scope of this paper, we do not want to comment on the matter. Notably though, precedents exist in other industries for granting antitrust exemptions on matters of public importance. If coordinated pausing is deemed sufficiently important for managing AI risks, exploring similar limited exemptions may be warranted.

Regardless of this, a pausing agreement would have other limitations. We are skeptical that frontier AI developers would be willing to enter into a legally binding pausing agreement. And even if they do, monitoring and enforcing the pause would still be left to the private sector with little or no public assurance. This would be problematic because we think that democratic institutions need to be involved if dangerous capabilities are in fact a serious threat to public safety and security \cite{Anderljung_et_al_2023, Shevlane_et_al_2023, Seger_et_al_2023}. It is also worth noting that in many jurisdictions it is not possible or at least very difficult to “force” a contracting party to comply. In principle, participating developers can still decide not to pause and pay the contractual fine (even though this will likely cause severe reputational damages).

\textbf{Recommendation.} To clarify the antitrust concern, developers may want to consult a specialized law firm to write a somewhat authoritative legal opinion on the topic. We also encourage legal scholars to analyze the question in detail \cite{Hua_Belfield_2021}. In general, we think that a pausing agreement would be better than voluntary pausing (Section \ref{3.1}), but it would still not be ideal. We would therefore prefer an audit agreement (Section \ref{3.3}) or pausing requirements (Section \ref{3.4}). In the meantime, we recommend that Anthropic, Google DeepMind, Microsoft, and OpenAI give the Frontier Model Forum the mandate to oversee their evaluation activities and, most importantly, membership in the Forum should require a pausing commitment.

\subsection{Mutual auditor}\label{3.3}

In the third version, a single auditor evaluates models of multiple developers who agree to pause if any model fails a set of evaluations.

\begin{figure}
    \centering
    \includegraphics[width=0.9\textwidth]{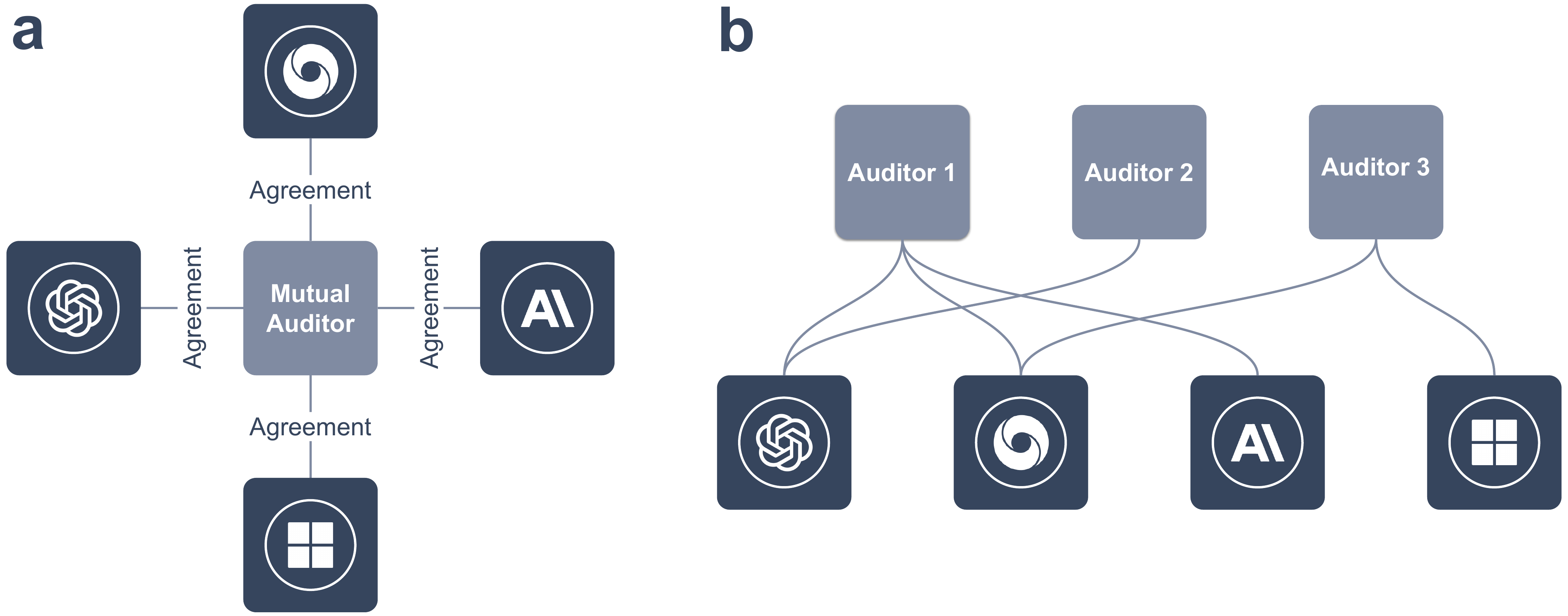}
    \caption{All participating developers commission the same auditor \textbf{(a)} or different developers commission different auditors \textbf{(b)}}
    \label{figure_3}
\end{figure}

\textbf{How it works.} All participating developers make an agreement with the same external auditor (Figure \ref{figure_3}a). They authorize the auditor to run dangerous capabilities evaluations on all frontier models they develop. Evaluations are developed and updated by the auditor with input from the developers. The developers commit to pause certain research and development activities if the auditor informs them that one of their models has failed a set of evaluations. They also give the auditor permission to notify other developers about the incident. Inversely, they commit to pause certain research and development activities if the auditor notifies them that a model from another developer has failed a set of evaluations. Finally, they commit to only resume the paused activities if the auditor gives them permission to do so. At the moment, ARC Evals seems to be the only organizations that would be able to serve the role of a mutual auditor \cite{ARC_Evals_2023b}. However, we suspect that more organizations will be set up in the future.

\textbf{Variation.} Instead of commissioning the same auditor, different developers could make agreements with different auditors (Figure \ref{figure_3}b). Auditors may be highly specialized organizations who run their own evaluations (e.g. ARC Evals and Apollo Research), or large audit firms without deep evaluation expertise (e.g. KPMG and Deloitte) who subcontract researchers or specialized organizations. However, any failed evaluation from any auditor would have to initiate a pause as described above. The auditor of the potentially dangerous model could either inform other auditors or other developers. If the auditors serve as licensed private regulators, this variation would be very close to the proposal of a “regulatory market” \cite{Hadfield_Clark_2023}. The value of multiple auditors is that it reduces the coordination effort required to agree on a mutual auditor. Developers have more choice, both with respect to evaluations and the auditors running them. It also allows developers to discover multiple failure modes, instead of just one standard set. On the flip side, it is harder to enforce a coordinated pause. For example, it will be difficult to agree on a danger threshold across different evaluations that different auditors run. In some cases, developers who are lagging behind might even want to trigger a pause to catch up.

\textbf{Benefits.} This version has three main benefits. First, the quality of the evaluations would be more consistent. The same actor would run the same evaluations, following the same process, using the same danger and safety thresholds. If the auditor accepts input from the wider AI safety community, their evaluations might actually represent the current state of the art, especially if they are routinely updated. Second, third-party evaluations tend to be less biased than internal evaluations. As a result, it is more likely that a pause will actually be imposed if necessary. Third, since the auditor would have access to the models, they can monitor compliance, at least to some extent.

\textbf{Limitations.} The following limitations seem most important to us. First, developers might be hesitant to give too much power to a single auditor, especially if the auditor has discretion and needs to make subjective judgments. Pausing certain research and development activities would have significant consequences for developers. They might lose millions or even billions of dollars in revenue, undermine their market position, and risk negative PR. They might only be willing to expose themselves to such risks, if they trust the auditor, the evaluations are sufficiently objective, and their main competitors also participate. But even this might not be enough. Second, in some cases, pausing might not be enough. The model that has failed the evaluations might already be so dangerous that simply pausing the training run might be an insufficient countermeasure. This might include cases where some sort of paradigm shift would be needed to avoid similar safety incidents. Third, some evaluations are similar to gain-of-function research. To see if a model has certain dangerous capabilities, the evaluator tries to elicit such behavior. Depending on the behavior, this type of evaluation might be extremely dangerous (e.g. power-seeking behavior). If such evaluations are conducted by an irresponsible actor, the measure might ultimately increase the risk.

\textbf{Recommendation.} This version seems particularly promising to us. It seems to be close to the sweet spot between desirability (i.e. it would be good from a societal perspective) and feasibility (i.e. there is a realistic chance that it would be implemented). Different stakeholders within and outside frontier AI developers should advocate for this option and policy makers should encourage it (e.g. in meetings with senior executives of frontier AI developers \cite{The_White_House_2023a}). This option should also be on the agenda of the upcoming global summit on AI safety \cite{HM_Government_2023}.

\subsection{Pausing requirements}\label{3.4}

In the fourth version, developers are legally required to run evaluations and pause if dangerous capabilities are discovered.

\textbf{How it works.} New laws or regulations require frontier AI developers to commission an external auditor to run dangerous capabilities evaluations on all frontier models. These laws also require developers to pause certain research and development activities and immediately notify a regulatory body whenever and each time one of their models fails a set of evaluations. This body, in turn, alerts other developers, asking them to also suspend similar activities. An independent investigation into the incident is then initiated by the auditor, in collaboration with the developer and under the regulator's supervision. Compliance is overseen by the regulator, who possesses investigative authority and can levy administrative fines. The decision to resume paused activities is made by the regulator, based on recommendations from the auditor.

\textbf{Variation.} Instead of creating new laws or regulations, regulatory bodies could try to use existing powers to enforce a pause. For example, the US Federal Trade Commission (FTC) has recently opened an investigation into OpenAI \cite{Zakrzewski_2023}. While the investigation focuses on potential violations of consumer protection laws, it seems plausible that the FTC or other regulators would also intervene in situations where it becomes publicly known that a model has failed a set of dangerous capabilities evaluations, creating incentives via legal liability. A detailed analysis of different powers of different regulatory bodies is beyond the scope of this paper.

Another way of implementing this model could be through regulatory markets \cite{Hadfield_Clark_2023}. This means that the government would set overall policy aims but rely on private regulators to determine the specific methods used for the intervention. In this case, developers would be legally required to purchase regulatory services from approved auditors. Auditors would be empowered to run evaluations, mandate pausing if triggers are met, oversee investigations, and approve resuming activities. The government provides ongoing oversight and can influence auditors through policy and incentives, but faces no pressure to gain state of the art technical expertise.

\textbf{Benefits.} The main benefit of this version is that it can ensure the highest levels of compliance. Depending on their precise powers, regulators can use various monitoring and enforcement measures to ensure that frontier AI developers actually run evaluations and pause if a model fails a set of evaluations \cite{Anderljung_et_al_2023}. It is also the only version where a democratically legitimated actor is involved in the pausing decision. If a model does in fact pose a serious threat to public safety and security, the government needs to be involved.

\textbf{Limitations.} Pausing requirements would also have a number of limitations. First, creating new laws and regulations takes time. However, many experts worry that frontier models might very soon be able to cause very severe harm (e.g. by enabling malicious actors to develop biological weapons). Second, frontier AI developers might lobby for weaker requirements. Although many developers actively support such requirements \cite{Amodei_2023, Altman_2023}, one should still be concerned of regulatory capture \cite{Anderljung_et_al_2023}. Third, regulators might be incentivized not to enforce the pausing requirements. The government might expect them to interpret their mandate in a laissez-faire, industry-friendly way (see e.g. \cite{UK_Department_for_Science_Innovation_Technology_2023}). Fourth, the introduction of pausing requirements would raise a number of further challenges. For example, it will be very difficult to define terms like “frontier AI model” and “dangerous capabilities” in a precise and future-proof way \cite{Anderljung_et_al_2023, Schuett_2023}.

\textbf{Recommendation.} We think that frontier AI developers should eventually be required by law to run evaluations and pause if dangerous capabilities are discovered \cite{Anderljung_et_al_2023, Shevlane_et_al_2023}. We highly recommend policymakers, above all the US and UK government, to seriously consider policy options along these lines. The recent announcement by the White House \cite{The_White_House_2023b}, which explicitly mentions “capability evaluations” and “dangerous capabilities”, is a promising step in this direction.

\section{Desirability}\label{4}

In this section, we discuss some of the benefits (Section \ref{4.1}) and potential harms (Section \ref{4.1}) of our proposed coordination scheme.

\subsection{Benefits}\label{4.1}

Coordinated pausing would have a number of benefits. But since it is a novel intervention, there is not yet any empirical evidence in support of these benefits. They are mainly based on abstract plausibility considerations.

\textbf{Preventing further scaling of dangerous models.} While running evaluations increases the chance that dangerous capabilities are discovered right after they emerge, our pausing scheme reduces the chance that developers further scale up such models. This is important because scaling up models with dangerous capabilities would likely make them even more dangerous. Without a pausing scheme, it seems plausible that at least some developers would continue scaling up their own models, even though a model with dangerous capabilities has been discovered.

\textbf{Preventing the deployment of dangerous models.} Pausing also reduces the risk that models with dangerous capabilities are deployed. Although models might already pose some risks before they are deployed (e.g. because they are used internally, leaked, or stolen), most risk scenarios require models to be deployed (i.e. made available to the public). While most developers would probably not deploy a model that has failed a set of dangerous capabilities evaluations, it seems plausible that other developers would continue deploying similar models. This would be bad because one might expect that similar capabilities will emerge in similar models. Put simply, pausing turns would-be catastrophes into warning shots.

\textbf{Buying more time for safety research.} Pausing creates more time for safety research. During the pause, safety researchers can study why a model has failed its evaluations, how to make the model safer, and what safety controls would be adequate (Section \ref{2.4}). They might also discover other safety issues. We think that buying more time for safety research may be one of the main benefits of our proposed pausing scheme. It seems plausible that safety research in this period is particularly valuable, mainly because it is possible to conduct empirical research on real models that pose real dangers. In the past, a lot of safety research was either theoretical or relied on toy models. The underlying principle—promoting risk-reducing technologies while delaying risk-increasing ones—has been referred to as “differential technological development” \cite{Bostrom_2002, Ord_2020, Sandbrink_et_al_2022}.

\textbf{Slowing down a race to the bottom.} Pausing might slow down a race to the bottom on safety. Commercial pressure might incentivize developers to cut corners on safety to get ahead of their competitors \cite{Armstrong_Bostrom_Shulman_2016, Naude_Dimitri_2020}. For example, after OpenAI released ChatGPT, Google famously announced it would “recalibrate” the level of risk it is willing to take \cite{Grant_2023}. If a developer gets an advantage by neglecting safety, others are incentivized to do the same. Otherwise, they might be left behind. However, during a pausing period, developers who have neglected their safety efforts would be able to catch up. This would at least temporarily stop a downward spiral.

\textbf{Shifting the Overton window.} Coordinated pausing might contribute to shifting the Overton window for other safety interventions, such as introducing strict domestic regulations on frontier models \cite{Anderljung_et_al_2023} or setting up new international institutions \cite{Ho_et_al_2023}. Every time a model fails a set of evaluations and participating developers pause, the incident would raise awareness of the dangers of frontier models. These “warning shots” would make other safety interventions increasingly politically feasible. We think that, although coordinated pausing may contribute to an Overton window shift, the effects of the scheme should not be overstated. In a world where some frontier models in fact fail dangerous capabilities evaluations, it seems likely that policymakers and the public would already be aware of the dangers and consider other interventions.

\textbf{Creating good incentives.} Coordinate pausing creates good incentives. Since pausing has a number of negative consequences, developers would likely want to avoid pauses. The most straightforward way a developer can avoid pauses is by ensuring that their own models and models of other developers pass evaluations. This provides an incentive to invest more in safety research and share insights with other developers.

\subsection{Potential harms}\label{4.2}

Below, we discuss ways in which our proposed coordination scheme might be harmful.

\textbf{Providing China with more time to catch up.} At the moment, Chinese AI companies seem to be behind their US competitors \cite{Ding_Xiao_2023, Toner_Xiao_Ding_2023}. However, one might worry that pausing frontier AI development in the US would give Chinese AI companies time to catch up.\footnote{A similar concern has been voiced in the regulatory debate \cite{Toner_Xiao_Ding_2023}.} Our best guess is that this concern is overblown. There are at least three reasons for this. First, we do not expect frontier AI developers in the US to pause for enough time for China to catch up in a meaningful way. Second, the US export controls on advanced computing and semiconductor manufacturing items \cite{US_Bureau_Industry_Security_2022, Shivakumar_et_al_2022} make it harder for Chinese AI companies to get access to cutting edge chips, which are necessary to train frontier models. This seems to be a meaningful constraint, even though Chinese firms have found some ways to evade the restrictions \cite{Fist_et_al_2023}. Third, Chinese AI companies have less incentives to develop frontier models, especially language models, mainly because they fear repercussions from the Chinese Communist Party (CCP).

\textbf{Providing a false sense of security.} One might worry that frontier AI developers and other stakeholders (e.g. regulators) rely too much on evaluations and coordinated pausing as their main intervention to reduce catastrophic risks from AI. This would be problematic if our proposed scheme alone is insufficient—which it probably is—and additional measures are not pursued. Pausing might buy negligible time for extra safety research, while providing a false sense of security to capabilities researchers. Again, our best guess is that the concern is overblown. We think it is unlikely that people would overly rely on the scheme. There seems to be an overall consensus among AI governance scholars and practitioners that there are no silver bullets and we need many different interventions (“defense in depth”). People would rather see it as yet another mechanism in a portfolio of mechanisms. For example, in a recent proposal for frontier AI regulation, dangerous capabilities evaluations would only inform a broader risk assessment \cite{Anderljung_et_al_2023}.

\textbf{Maintaining market position.} It is possible that, if this intervention is implemented, developers ahead in capabilities would have incentives to dishonestly trigger pausing periods to delay the progress of their competitors. This risks distorting the entire scheme by transforming pauses into a mechanism for suppressing competition rather than promoting safety. To mitigate this, it is crucial that evaluations are conducted transparently (even if they are developed in-house and on a voluntary basis), so that other researchers can ensure the evaluations serve their intended safety purpose. Whistleblowing schemes could add such a layer of transparency, making it riskier for lab leadership to manipulate the intervention. Given that such manipulation would likely require coordination across multiple teams, the presence of a whistleblowing mechanism would reduce the likelihood of internal trust sufficient for such a scheme.

\textbf{Discontinuous scaling.} Rapid AI capability advancements could occur post-pause. If developers continue to make (even restricted) algorithmic improvements while paused, they could create a sudden leap in capabilities once the pause is lifted. For example, if developers are restricted to working on smaller models during a pause, they might focus on fine-tuning and developing new techniques. When they are allowed to use more computing power again, these improvements could combine to create a big jump in capabilities, catching regulators off guard. This could even lead to a “hard takeoff”, where AI capabilities advance very quickly \cite{Yudkowsky_2013}. To prevent this, it's crucial that the pause restrictions are designed to actually slow down capabilities research, not just limit deployment.

\textbf{“Wolf cries”.} While the open letter “Pause Giant AI Experiments” \cite{FLI_2023} has received some support \cite{Bengio_2023}, it has also been criticized \cite{Ienca_2023, Paul_2023}. In general, it has likely contributed to push backs against the concern that future AI systems might cause catastrophic or even existential risks. Skeptics see a discrepancy between current capabilities and warnings of imminent threats. One might worry that if capabilities become more dangerous, people will take justified warnings less seriously. This situation is similar to the fable of the boy who cried wolf.\footnote{A shepherd boy repeatedly fools villagers into thinking a wolf is attacking his town's flock. When an actual wolf appears and the boy calls for help, the villagers believe that it is another false alarm, and the sheep are eaten by the wolf.} Our proposed coordination scheme might make this scenario more likely. We wish to emphasize that current warnings might very well be justified, not because existing models already pose catastrophic risks, but because we need to be prepared for scenarios in which the next generations of models do.

In this section, we have discussed the main benefits and potential harms of coordinated pausing. We conclude that coordinated pausing is a promising mechanism for tackling emerging risks from frontier AI models. But could it actually be implemented?

\section{Feasibility}\label{5}

This section discusses the feasibility of our proposed coordination scheme. The following factors seem most important, that is, we expect the intervention to fail if these factors prove to be insurmountable.

\textbf{Violation of US and EU antitrust law.} One concern is that coordination between AI developers could violate antitrust laws in the EU and US. In the European Union, Article 101(1) of the Treaty on the Functioning of the European Union prohibits and nullifies agreements between companies that have the effect of restricting competition. A coordinated commitment to pause and any communication between developers about pausing plans could potentially violate this law.

However, there are ways developers may be able to avoid this issue. For example, they can make independent commitments to pause without discussing them with each other. This avoids any explicit agreement or \emph{concurrence of wills} between competitors. Similarly, when communicating about dangerous capabilities discoveries, developers can avoid explicitly mentioning plans to pause or encouraging others to do so as well. Using third parties like independent auditors or regulators as intermediaries for sharing information may also help mitigate these concerns. For example, a mutual auditor can notify developers of failed evaluations without the developers communicating directly. Finally, developers can avoid sharing commercially sensitive information about models with one another, to the extent it is possible.

In the United States, Section 1 of the Sherman Antitrust Act lays out the relevant law on this matter, and  similarly prohibits conduct that unreasonably restrains trade in a way that is harmful for competition. In addition to the measures mentioned above, it is also crucial that any coordination scheme between AI developers in America does not include explicit restrictions on price or output that position developers to profit. Retaliatory actions against non-participating developers should also be avoided. Additionally, it is worth noting that US courts may not accept a defense of coordination schemes based on public policy merits if they are found to be anticompetitive. Consulting with regulatory bodies like the US Federal Trade Commission, the US Department of Justice, and the EU Competition and Markets Authority may be an important strategy to ensure compliance with antitrust law.

\textbf{Enforcement concerns.} Enforcing compliance with pausing commitments could also be challenging. This is especially the case for voluntary schemes that lack formal oversight. While public reputation provides some incentive to comply, it has limitations as an enforcement tool. For instance, the strength of public pressure could fade over time if pausing is rarely triggered, as media attention shifts to other issues. Developers might also try to strategically influence public discourse to portray the intervention negatively and reduce compliance pressure. Similarly, if highly profitable models emerge during pause periods, commercial incentives could override reputational concerns about violating commitments. Thus, while voluntary compliance is worth pursuing, robust legal authorities and enforcement tools would likely be needed to ensure developers adhere to pausing in impactful cases. Despite this, there may be some lower-cost enforcement options to consider. For example, cloud computing providers could be pressured into updating their terms of service to contractually enforce pauses by restricting access to resources during pause periods. Yet, overall, relying solely on public pressure will provide limited assurance.

Enforcing compliance through legal agreements with auditors also poses challenges. As previously mentioned, it is often challenging to force contracting parties to actually fulfill their obligations, rather than just pay damages for breaking the contract. Developers could choose to breach a pausing agreement and accept the financial consequences instead. However, structuring agreements so penalties are sufficiently large and tied to revenue could make violations prohibitively expensive, helping disincentivize non-compliance. But ultimately, a multipronged approach combining reputational incentives, contractual leverage, ongoing scrutiny, and collaborative partnerships between stakeholders will likely be needed to reliably enforce adherence.

\textbf{Model verification concerns.} Another potential obstacle is ensuring that the systems evaluated during training are the same systems that are eventually deployed. For instance, OpenAI's GPT-4 system card revealed the audited version of GPT-4 differed from the deployed model \cite{OpenAI_2023a}. Ensuring the integrity of this matching is critical, as deploying systems with capabilities differing from those assessed during development could enable developers to bypass the intervention and deploy unsafe models.

Several potential solutions exist: Auditors could check if the distribution of outputs on a secret benchmark dataset match between the audited and deployed versions. Hashing the trained model weights and having compute providers verify the hashes match is another option. Watermarking models in a way that is sensitive to fine-tuning, then checking the watermark pre and post-deployment could also work. In the future, requiring “signed” models approved by auditors may be possible if hardware only runs approved models. However, even if audited and deployed versions can be matched initially, models are often continuously updated after deployment, potentially developing new dangers. Requiring re-evaluations before modifications or limiting live tuning may help, but could also hamper capabilities. Ongoing monitoring of deployed systems for emerging issues will likely be needed.

\textbf{Goodharting.} Developers may try to manipulate evaluations in order to avoid triggering pauses, i.e. training their systems in ways that ensure they will reliably pass model evaluations, even if they are dangerous. However, certain evaluation design choices could help mitigate this risk. For instance, using private benchmark datasets that are not included in the training data for future models may make it harder for developers to unfairly optimize performance. Additionally, keeping some aspects of evaluation methodologies opaque could increase the difficulty of gaming them. Furthermore, lengthy dynamic evaluations involving humans in the loop at multiple stages would limit the ability of developers to rapidly iterate and overfit to the assessment. If evaluations are designed to be robust and multifaceted, requiring many iterations to reverse engineer and bypass, the risks of Goodharting may be reduced.

\textbf{No difference between safety and capabilities research.} A key premise of our proposed intervention is that it provides time for safety research to progress separately from capabilities development. Yet some safety techniques like Reinforcement Learning from Human Feedback also improve capabilities. This means that developers conducting safety research during pauses may inadvertently continue to advance capabilities, even if direct capabilities work is paused. However, this risk can be acknowledged and mitigated. Developers can conduct safety research under intentionally constrained conditions during pauses, limiting training data, compute resources, etc. to slow capability gains. Additionally, thresholds can be set for acceptable capability gains from safety work during pauses \cite{Hendrycks_Mazeika_Woodside_2023}. Therefore, even if safety and capabilities research are intertwined, it may be possible to maintain meaningful differentiated progress with deliberate effort and constraint.

The following factors seem less critical. While we do not expect our intervention to fail if progress is not made on these factors, we expect the feasibility of the intervention to drop considerably.

\textbf{No consensus on model evaluations.} A lack of consensus on the appropriate model evaluations to implement could also become an obstacle to the intervention’s feasibility. On the one hand, allowing for a diversity of assessments might increase the chances of detecting varied failure modes that a single standardized set of evaluations would miss. On the other, too much inconsistency could also reduce the likelihood that any given issue is caught across all developers. Furthermore, if participating developers implement wildly varying or self-serving evaluations, dangerous systems may slip through undetected, limiting the value of coordination.

However, some flexibility could still improve the status quo, provided leading developers conduct rigorous evaluations and do so responsibly. Moreover, reaching perfect agreement creates additional complexity when trying to convince developers to commit to pausing. Allowing some room for disagreement on precise metrics may increase willingness to align on the foundational intervention, if not every detail.

\textbf{Dissuasion from investors.} An additional worry is that investors could threaten to withdraw funding, dissuading developers from making pausing commitments that might slow research progress. Our sense is that this concern is surmountable. If leading developers coordinate around pausing, investors may have little choice but to continue funding them or accept lower returns from less capable labs. This dynamic persists as long as developers have multiple competing funding options, providing leverage. Additionally, both investors and AI developers are likely interested in appearing socially responsible. As long as pausing does not completely preclude promising research, the same incentives that persuade AI developers to participate should also keep investors onboard.

\textbf{IP concerns.} Another potential obstacle is that developers may be reluctant to allow the level of external auditing this intervention requires due to concerns over intellectual property and confidentiality. Underpinning these concerns could be the idea that providing auditors structured access to models, such as through an API, may not be sufficient \cite{Bucknall_et_al_forthcoming}. Yet, it is possible that existing measures from other industries can alleviate this problem. For instance, developers can implement access controls like air-gapped evaluation rooms for auditors to view model weights in, a common information disclosure strategy within governments. 

IP concerns may also extend beyond simple information disclosure issues. For instance, AI developers might be concerned that auditors will be unable to adequately prevent sensitive information from being stolen by third parties. Securing lab information is already an extraordinary challenge because of the sheer size of the attack surface and the incredible influence of potential adversaries \cite{Ladish_Heim_2022}. Auditors are unlikely to have the same level of defensive resources as top labs and, as a result, may represent an additional layer of vulnerability for labs. However, it may be worth noting that upcoming regulations like the EU AI Act, as well as recent commitments by frontier developers visiting the White House, may necessitate external audits regardless, suggesting developers already have strong incentives to find solutions here.

In this section, we have discussed whether our proposed coordination scheme is actually feasible. Overall, we are cautiously optimistic that the practical obstacles hindering the intervention are surmountable. But successfully implementing such coordination will require care, foresight, and continued research.

\section{Conclusion}\label{6}

This paper has proposed an evaluation-based coordination scheme for situations in which frontier AI developers discover that their models have certain dangerous capabilities (Section \ref{2}). Such a scheme could rely on public pressure, a pausing agreement, a mutual auditor, or legal requirements (Section \ref{3}). The paper has also discussed the desirability and feasibility of the proposed scheme (Section \ref{4} and \ref{5}). We concluded that coordinated pausing is a promising mechanism for tackling emerging risks from frontier AI models. However, a number of practical and legal obstacles need to be overcome.

\textbf{Questions for further research.} This paper has left many questions unanswered and more research is urgently needed. The following six areas seem particularly important:

\begin{itemize}[leftmargin=2em]
    \item \textbf{Dangerous capability evaluations.} The most obvious bottleneck of the proposed coordination scheme is a lack of reliable evaluations for dangerous capabilities. We are only aware of ready-to-use evaluations for power-seeking behavior \cite{ARC_Evals_2023a, Kinniment_et_al_2023}, though we expect that some developers have internal evaluations that they do not share publicly. Evaluations for other dangerous capabilities discussed in the literature \cite{Shevlane_et_al_2023} do not yet exist, even though there are efforts to create them. We strongly encourage researchers and practitioners to create new evaluations and scrutinize existing ones.
    \item \textbf{Safety thresholds.} Defining danger thresholds (when should frontier AI developers pause?) and safety thresholds (when can they resume their paused activities?) are still open questions which require more research.
    \item \textbf{Model similarity.} We have skipped the questions of which models should be considered “similar” to those which have failed evaluations, and which research should be considered “related” to that which led to the failed evaluations. This raises a number of thorny questions.
    \item \textbf{Developer buy-in.} We encourage more work that investigates ways in which developers can be incentivized to run evaluations and to participate in the proposed coordination scheme. For example, this might involve frontier AI regulation \cite{Anderljung_et_al_2023} or advocacy aimed at increasing public pressure on developers.
    \item \textbf{Legal considerations.} There is some uncertainty over whether some versions of this intervention might violate antitrust law. It would be valuable to know to what extent these concerns are justified. If antitrust law is in fact a meaningful constraint, one could investigate options for a narrow safe harbor for coordinated pausing \cite{Muehlhauser_2023}. Section 708 of the US Defense Production Act (DPA), which can shield companies cooperating under the DPA from antitrust liability, might be a promising tool. The tool has already been used during the COVID-19 pandemic \cite{The_White_House_2020, Lobert_2020}.
    \item \textbf{Internal response policies.} This paper has focused on a collective solution, i.e. what multiple developers should do if one of them discovers a model with sufficiently dangerous capabilities. A related question that warrants further attention is what exactly a single developer who discovers these capabilities should do. Recently, Anthropic laid out its Responsible Scaling Policies: internal safety measures they plan to implement before scaling up models \cite{Anthropic_2023c}. Similar work has been published by ARC Evals \cite{ARC_Evals_2023c}. These policies are based on evaluations and include commitments to pause training and deployment for models that fail Anthropic's assessments. Developing similar evaluation-triggered internal response protocols is an important area for future work. These policies should also specify the conditions under which developers can resume any paused activities.
\end{itemize}

In their latest update, ARC Evals concluded that the versions of Claude and GPT-4 they tested did not have sufficiently dangerous capabilities, but their outlook was concerning: “for systems more capable than Claude and GPT-4, we are now at the point where we need to check carefully that new models do not have sufficient capabilities to replicate autonomously or cause catastrophic harm—it’s no longer obvious that they won’t be able to” \cite{ARC_Evals_2023a}. We urge policymakers, researchers, and practitioners to take this warning seriously. We need to be prepared for a world in which Claude 3 or GPT-5 fail their evaluations. We believe that coordinated pausing needs to be part of any solution.

\section*{Acknowledgements}

We are grateful for valuable feedback and suggestions from Akash Wasil, Alan Chan, Andrea Miotti, Ben Garfinkel, Daniel Kokotajlo, Hjalmar Wijk, Holden Karnofsky, Jack Clark, Lennart Heim, Lukas Gloor, Markus Anderljung, Nate Soares, Nathan Calvin, Noam Kolt, Noemi Dreksler, Francis Rhys Ward, Vivian Dong, and Zvi Mowshowitz (in alphabetical order). All errors are our own.

\bibliographystyle{abbrv}
\bibliography{ms}

\end{document}